\documentstyle[11pt,aaspptwo,flushrt,tighten]{article} 

\newcommand{\ltsima} {$\; \buildrel < \over \sim \;$}
\newcommand{\gtsima} {$\; \buildrel > \over \sim \;$}
\newcommand{\lta} {\lower.5ex\hbox{\ltsima}}
\newcommand{\gta} {\lower.5ex\hbox{\gtsima}}

\begin{document}

\title{X-RAY VARIABILITY AND CORRELATIONS\\
IN THE TWO PHASE DISK--CORONA MODEL FOR SEYFERT GALAXIES }

\author{Francesco Haardt\altaffilmark{1,2}}
\affil{Department of Astronomy \& Astrophysics, Institute of
Theoretical Physics, G\"oteborg University \& Chalmers University
of Technology, 412 96 G\"oteborg, Sweden} 

\author{Laura Maraschi and Gabriele Ghisellini\altaffilmark{1}} 
\affil{Osservatorio Astronomico di Brera/Merate, 20100 Milano, Italy}

\altaffiltext{1}{ITP/UCSB, Santa Barbara, CA, USA}
\altaffiltext{2}{Present address: Universit\'a degli Studi di 
Milano, Italy}

\begin{abstract}
We discuss in detail the broad band X--ray variability expected in Seyfert
galaxies in the framework of a Compton cooled corona model, computing
explicitly spectral indices in the 2--10 keV range and fluxes in the medium
(2--10 keV), hard (30--100 keV) and soft (0.1--0.4, 0.9--2 keV) X-ray bands, for
different hypotheses concerning variability "modes". In all cases  the soft 
photons responsible for the cooling are assumed to be proportionate to the
Comptonized flux (fixed geometry), and the plane parallel limit is adopted.
Variations in the optical depth $\tau$ of the corona cause  spectral changes in
the Comptonized emission in the sense that the spectrum steepens and the
temperature decreases with increasing $\tau$. If the corona is pair dominated,
$\tau$ is determined by the compactness (proportional to luminosity for fixed
size). This yields a definite relation between spectral shape and  intensity
implying a variation in intensity of a factor 10 for a change of 0.2 in the
2--10 keV spectral index. For low pair density coronae, no definite prediction
is possible. If
for instance $\tau$ varies while the luminosity remains approximately constant,
the Compton spectrum shows a pivoting behaviour around $\simeq$ 10 keV. In
both cases,  the coronal temperature decreases for increasing spectral index.
In the absence of substantial changes in the structure of the corona, the soft
thermal component  correlates with the medium--hard Comptonized component.
Although the correlation must hold globally, that is for the integrated power
of the two components, the observed intensity in the soft X-ray band is
extremely sensitive to the temperature of the thermal component which may vary
as a result of variations in luminosity and/or area of the reprocessing
region(s). Examples are computed using the ROSAT response matrix, showing that
for $kT\lta 60$ eV the spectrum in the ROSAT range softens with 
increasing intensity while
the opposite is true for larger temperatures.  Current observations of NGC 5548
and Mrk 766 showing different behaviours can therefore be reconciled with the
model. The spectral behaviour of Mrk 766 strongly suggests that the coronal
plasma in this source is not pair dominated. 
The possibility of constraining models through simultaneous observations in the
soft medium and hard X-ray ranges is discussed. 
\end{abstract}


\twocolumn 

\section{Introduction} 

The observations of OSSE on the Compton Gamma Ray Observatory showing a break
or an exponential cut--off in the high energy spectrum of some Seyfert galaxies
(Maisack et al. 1993, Johnson et al. 1993, Madejski et al. 1995) coupled with
the X-ray observations of GINGA, have had a strong impact on our theoretical
understanding of the X--ray and $\gamma-$ray emission processes in these
objects. Though they do not rule out pure non--thermal $e^{\pm}$ pair models
(Zdziarski, Lightman \& Maciolek-Niedzwiecki 1993), thermal or quasi--thermal
Comptonization models are favoured (e.g. Sunyaev \& Titarchuck 1980; Haardt \&
Maraschi 1991 and 1993, hereinafter HM91 and HM93; Ghisellini, Haardt \& Fabian
1993; Titarchuck \& Mastichiadis 1994; Zdziarski et al. 1994; Zdziarski et al.
1995; see Svensson 1996 for a recent review). 

The origin of the Comptonizing electrons is at present unknown. They could be
related to an accretion disk corona heated by magnetic dissipation processes
and/or they may be electron positron pairs produced by photon photon collisions
(see e.g. Fabian 1994 and references therein). The electron temperature can be
constrained by a thermal balance equation, where the heating rate is the total
(direct plus reprocessed) luminosity and the cooling mechanism is
Comptonization of soft photons. 

The hot corona is probably coupled to a cooler optically thick gas layer
(presumably the accretion disk itself), which has three major roles: i) it
reprocesses a fixed fraction of the Comptonized radiation into a soft blackbody
component;  ii) a fixed fraction of the reprocessed blackbody photons crosses
the hot region acting as seed photons for the Comptonization process; iii) it
reflects medium energy X-rays giving rise to the so called reflection component
and Fe line emission. This picture (HM91, HM93) is suggested by the presence of
the Compton reflection hump (Pounds et al. 1990) and by the relatively small
dispersion in the distribution of spectral indices of Seyfert I galaxies
(Nandra \& Pounds 1994). In fact, if  X-ray reprocessing contributes soft
photons proportionately  to the medium-hard X-ray luminosity, the 'feed-back'
effect keeps a quasi-constant X-ray spectral shape, weakly dependent on the
source luminosity. The broad band spectra derived from this model are
consistent with the average Seyfert spectrum derived from OSSE observations 
(Zdziarski et al. 1995; Gondek et al. 1996).

It is fair to point out that the best studied object, NGC 4151, shows 
(at least in low state) an unusually flat X-ray spectral index
($\alpha\simeq 0.5$) and a low temperature ($kT\simeq 60$ keV) implying a
Comptonized--to--thermal luminosity ratio $L_c/L_s \gta 10$, far from the limit
$L_c/L_s \simeq 2$ obtained in a plane parallel geometry.  This may be
accounted for by a special geometry, for instance
if the high energy emission originates in localized non planar active blobs
as in the model proposed by Haardt, Maraschi \& Ghisellini 1994 (hereinafter
HMG). Alternative models involving a mirror reflection of the observed
radiation (Poutanen et al. 1996) or reprocessing in clouds along the line of
sight (Zdziarski \& Magdziarz 1996) may also apply for this special object.  

In Comptonization models the spectral shape in the X--ray range is mainly
determined by the combination of the temperature $\Theta=kT/mc^2$, and the
optical depth of the scattering electrons $\tau$, while the cut off energy is
related essentially to $\Theta$. Simultaneous measurements of the spectral
index $\alpha$ and of the cut off energy can therefore determine 
the physical parameters of the Comptonizing region\footnote{The exact relation 
between $\alpha$ and $\tau$ depends on the geometry of the scattering region.  
The optical depth can therefore be deducted from observations only "modulo 
geometry". On the other hand fitting the data assuming a 
simple homogeneous sphere gives a value of $\tau$ that is also
a good estimate of the average (over all directions) scattering optical 
depth for any other possible geometry.}.

``Spectral dynamics'',
i.e., the variation of the spectral shape with intensity is an additional basic
diagnostic tool (see, e.g., Netzer, Turner \& George 1994) which can test the
model and provide insight into the way the corona is  heated and cooled. Some
of the questions that could be answered are: is the optical depth of the corona
dominated by $e^+e^-$ pairs? How do the optical depth and electron temperature
vary with luminosity? Are the reprocessed components proportionate to the
Comptonized one during variations? For instance a careful analysis of  EXOSAT
spectra of Seyfert galaxies, showed that the observed spectral variability is
difficult to reconcile with non--thermal pair models (Grandi, Done \& Urry,
1994) . 

Another important prediction of the model is the presence of a thermal 
component which should be proportionate to the Comptonized one in the limit of
a constant geometry. Its temperature however is not completely constrained  by
the model and could range from UV to soft X-rays. The presence of a "soft
excess" in the X--ray spectra of AGN has been detected by different experiments
(EXOSAT, ROSAT and more recently ASCA)  (e.g. Arnaud et al. 1985; Pounds et al.
1986; Wilkes \& Elvis 1987; Turner \& Pounds 1989; Elvis, Wilkes \& Mc Dowell
1991; Comastri et al. 1991; Saxton et al. 1993; Walter \& Fink 1993; Turner,
George \& Mushotzky 1993; Brandt et al. 1994; Done et. al 1995).  It has been
interpreted in some cases as a genuine thermal component possibly 
due to reprocessing 
(e.g. Leighly et al. 1996), 
or as the high energy tail of the UV bump (e.g. Walter \& Fink 1993), while 
in others as an artefact of the residuals due to 
incomplete absorption (warm absorber, see e.g. Done et al. 1995). 
Instruments with improved energy resolution in the soft band are
needed to clarify the interpretation of
the "soft excess"  which may even involve more than one component.
However a distinctive point between different models is the predicted
correlation between the soft and medium X-ray intensities. On this aspect
useful data  obtained from extensive observations of  some Seyfert galaxies
with ROSAT and/or ASCA already exist and more will surely become available in
the near future. 

We are therefore motivated to examine in detail the  broad band spectral
variability expected in the accretion disk with hot corona model developed in
previous papers (HM91, HM93 and HMG). We will concentrate on  the case of
extragalactic objects such as Seyfert I galaxies. The issue is complicated for
several reasons. A quantitative analysis requires spectral simulations since
"second order effects" are important (see \S 3) . However the main uncertainty
is our poor knowledge of the physics of the corona. For instance do the active
coronal regions vary in size and/or height above the reprocessing disk? And how
is the heating rate related to the optical depth? 

In the following we chose the simple case in which the active and reprocessing
region(s) can vary in area but maintain a fixed ratio between Comptonized and
soft cooling photons. In particular we choose the limit of a plane-parallel
geometry characterized by a Comptonized--to--soft ratio $L_c/L_s\simeq 2$. This
is not inconsistent with an inhomogeneous picture where each active region can
be approximated as plane parallel. With these restrictions we compute
explicitly the expected relations between "observable quantities", e.g. the
variation 
of the 2--10 keV spectral index (including reflection) with
temperature and intensity,  
the variation of the soft intensity "S"(0.1--2 keV) with
medium "M" (2--10 keV) and hard "H" ($E\gta 30$ keV) intensity. 

The plan of the paper is as follows. In \S 2 we describe the model and the
methods used for the computation of the spectra and for the derivation of the
"observable" quantities. Since the key parameter driving spectral variability in the 
medium band is $\tau$, while spectral variability in the soft band is 
essentially related to $T_{BB}$, we found convenient to discuss relations 
between spectral and flux variability in the two bands separately. 
For this reason, in \S 3 we deal with
the expected spectral variability
in the medium and hard X--ray bands, using the 2--10 keV spectral index, the
2--10 keV intensity, the temperature of the hot corona and the 30--100 keV
intensity, while in \S 4 we compute the intensity and spectral variations of the
soft reprocessed blackbody component for different variations of the system
parameters. The results are converted into count rates and hardness ratios in
the 0.1--0.4 and 0.9--2 keV bands using the ROSAT response matrix and a standard
neutral hydrogen column. We discuss the expected correlations between
variability in the soft and medium energy bands and compare the results to
available  observations in \S 5. An exhaustive summary and final 
conclusions are presented in \S 6.

\section{The Disk--Corona Model} 

We follow a scenario originally proposed by Liang (1979), and revived
by HM91 and HM93, in which the accretion flow is illuminated by the
hard X--ray source (as the observed reflection spectrum indicates). A large
fraction (about $80-90$ per cent) of the incident power is expected to be
thermalized in the disk and re--emitted as thermal reprocessed radiation, which
is in turn Comptonized in the corona producing the X--rays. If the reprocessed
radiation energy density dominates over the local disk thermal emission, the
Compton $y$ parameter is forced to be $\simeq 1$ independently of the coronal
scattering opacity. This in turn ensures that the Comptonized radiation
exhibits a spectral index close to 1. One possible configuration which
satisfies the above condition is a plane parallel corona above an accretion
disk. In this case the condition $y\simeq 1$ is achieved only if the {\it
entire} available gravitational power is released in the hot corona. In this
case the {\it reprocessed thermalized} radiation coincides with the UV 
bump, and the {\it reprocessed reflected} radiation forms the 30 keV
Compton hump. 

The same condition can be satisfied if there are several such smaller regions
above the disk, rather than a single smooth corona (HMG). In the latter case
only a fraction of the available energy needs to be released in the hot corona. 
However being localized in hot spots, 
the reprocessed thermalized radiation can still
dominate the cooling.  The underlying accretion disk can contribute to most or
at least part of the total luminosity in the UV band as in the standard
Shakura--Sunyaev disk model (Shakura \& Sunyaev 1973). As in the uniform corona
case, the {\it reprocessed reflected} radiation forms the 30--keV Compton hump,
while the {\it reprocessed thermalized} radiation is probably emitted in hot
spots  in the EUV--soft X--ray band (HMG). Therefore the emission in the
EUV--soft X--ray band should be closely connected to the hard X--ray radiation.

In the present paper we assume that the cooling time of the electrons in the
corona is shorter than the time scale of variation of the heating rate, so that
we can consider spectral and flux variability as a succession of stationary
states. The Comptonized spectra are computed in plane parallel geometry for
different inclination angles $\vartheta$. A quadrature scheme was implemented
to solve the radiation field for successive scattering orders, taking into
account anisotropic Compton emission for the first scattering distribution
(Haardt 1993). The single scattering photon distribution is computed by means
of the fully relativistic kernel for isotropic unpolarized radiation (Coppi \&
Blandford 1990). This ensures that the exponential roll over of the spectrum is
correctly modeled as we tested against Montecarlo calculations. The 
present numerical spectra are therefore in much closer agreement
with recent detailed calculations by Poutanen \& Svensson (1996) than
 the approximated treatment of HM93. An extensive 
discussion of the method we used can be found in Haardt (1994). 

The downward Compton flux (i.e. the X--ray flux leaving the corona with an
angle $\vartheta>\pi/2$) is averaged over the solid angle, and then reflected
by an infinite neutral layer (which mimics the accretion disk) with solar metal
abundances. The angular distribution of the reflected spectrum is computed in
the approximated way described in Ghisellini, Haardt \& Matt (1994) (see
Burigana 1995 and Magdziarz \& Zdziarski 1995 for detailed calculations). The
angular dependent reflected spectrum is then Comptonized in the corona and
added to the primary continuum. The "measurable" spectral index in the [2--10]
keV band is obtained fitting with a power law to the composite spectrum
disregarding the goodness of the fit. The photon flux (i.e. number of photons
per unit time per unit surface) in interesting bands is readily obtained by
integration over energy. 

Concerning the soft photon input, we consider cases in which the reprocessed
radiation is isothermal and has a temperature $T_{BB}$ comprised between 10 and
300 eV. This scenario can be regarded as a uniform corona located above the
very inner part of the accretion disk, or alternatively as a blobby structure
where the overall X--ray emission is dominated by few blobs with similar
properties. 

With our approximations the interesting quantities are computed with good 
accuracy in a quick way. A more detailed treatment of the full radiative
transfer problem in the same context, including free--free radiation, double
Compton scattering, detailed pair balance and polarized scattering, can be
found in Poutanen \& Svensson (1996). 

\section{Variability in the Medium and Hard X--Ray Domains} 

\subsection{The $\alpha-\tau$ Relation}

Since the model constrains the Compton parameter to be almost constant, one
expects that the spectral index of the Comptonized spectrum  is fairly
constant too, and does not depend on the three  parameters of the
system (i.e. $\tau$, $\Theta$ and $T_{BB}$). 

The above argument is however valid only as a first order approximation,
in particular in the case of a disk--corona system, where the scattering anisotropy 
is important and  when we
consider the actually  observed spectrum which includes the reflection hump.  
We therefore computed the spectral index of the {\it
total} spectrum in the [2--10] keV range $\alpha_{[2-10]}$ 
as described in the previous
section for different values of the coronal parameters and of the 
inclination angle. Comptonization of the reflected spectrum is also 
included.

The results are shown in Fig. 1 and Fig. 2 where $\alpha_{[2-10]}$ is plotted
vs. the optical depth and the temperature of the corona respectively, for two
values of the inclination angle and two values of the  temperature of the soft
photons. In order to show the effect of reflection on the total spectrum, 
in Fig. 3 we plotted, as a function of $\tau$, the difference between the  spectral index of the intrinsic continuum alone and that of the total spectrum (continuum+reflection). 

The optical depth was varied between 0.1 and 1, corresponding 
to an equilibrium temperature of the hot corona in the range 30--300
keV, as suggested by observations (Maisak et al. 1993, Johnson et al. 1993,
Madejski et al. 1995). 

The important points to note are the following: 

\begin{enumerate}
\item{}Changes  in $\tau$ give rise to significant spectral variability
despite the fact that the ratio $L_c/L_s$ is constant in the model. In the
plane parallel limit considered here $\alpha_{[2-10]}$ is found to be in the
range [0.4--1.4] for $\tau$ varying between 0.1 and 1. 
\item{} Steeper spectra are produced as $\tau$ increases approaching unity. 
\item{} An increase of the spectral index is accompanied by a  decrease of
the coronal temperature, i.e. {\it spectral index and coronal temperature are
anticorrelated}. The actual values of $\alpha$ and $kT$ in Fig 2 refer to
$L_c/L_s \simeq 2$, however the anticorrelation between  spectral index and the
electron temperature is a general feature common to models based on Compton
cooling with fixed $L_c/L_s$.  For higher values of $L_c/L_s$  a similar curve
would be obtained, but with values of $\alpha$  shifted downwards. As shown in
Fig. 2, the dispersion of the $\alpha_{[2-10]}$ vs. $kT$ relation is larger at
higher temperature. 
\end{enumerate}

A physical understanding of these results involves several effects: 

\begin{enumerate}
\item{} for any given $\tau$ the temperature has to adjust to maintain an
almost fixed ratio $L_c/L_s$, and for increasing $\tau$ the temperature
decreases. The resulting average spectral index (i.e. the spectral index
averaged over the total solid angle) is, for the plane parallel geometry
adopted here, slightly larger than 1 for any value of $\tau$ (HM93).
Comptonization theory implies that, in order to keep a constant ratio
$L_c/L_s$, a spectral index {\it steeper than 1} must increase with decreasing
$\Theta$ (i.e. increasing $\tau$). This can be understood noticing that the
extrapolation to low energy of the Compton power law must intercept the peak of
the {\it injected} black body soft photon distribution. Since for $\alpha>1$
the integral over energy of the power law 
spectrum mainly depends on the lower integration
limit $E_1 \propto (16\Theta^2+4\Theta+1)kT_{BB}$, a shift to lower energy of
$E_1$ (i.e. a decrease of $\Theta$) keeping a constant $\alpha$ would result in
a larger ratio $L_c/L_s$. Therefore, in order to keep the ratio constant, 
the spectral index is forced to steepen
(for the same reason, if $\alpha<1$, a reduction of $\Theta$ implies a further
{\it flattening}, rather than a steepening, of $\alpha$). The steepening of
$\alpha$ halts when a further decrease of $\Theta$ does not cause a shift of
$E_1$ to lower energy, i.e. when $16\Theta^2+4\Theta \ll 1$, say 
$16\Theta^2+4\Theta\lta 0.5$. For even smaller
temperatures the spectral index must {\it flatten} (and hence there is a
maximum in the $\alpha$ vs. $\Theta$ curve) since, though $E_1$ is constant,
the upper integration limit ($\propto \Theta$) decreases anyway. In terms of
$\tau$, since in our adopted geometry the Compton parameter $y\simeq 
(16\Theta^2+4\Theta)
\tau$ is $\simeq 0.6$ (HM93, Poutanen \& Svensson 1996), 
the "turning point" occurs for $\tau \simeq 1$. 
Note that for $\alpha<1$ ($L_c/L_s \gta 2$) there is a 
monotonic flattening of $\alpha$ as $\Theta$ decreases. The effect discussed
here is independent of the particular geometry adopted. In brief, a fixed
$L_c/L_s$ ratio does not imply an exactly constant spectral index [e.g. see
Fig.2 in Ghisellini \& Haardt (1994; hereinafter GH94) for the simplest
possible geometry, i.e. a homogeneous spherical source, where this effect is
present];\par 
\item{} the outgoing Compton spectrum in plane parallel geometry  is a
function of the emission angle, because of the Compton rocket effect (Haardt
1993). Flatter spectra are seen for lower inclinations, and the importance of
the anisotropic Compton emission decreases with decreasing coronal
temperatures. Anisotropic Compton emission 
is the main cause of the reduction of $\alpha_{[2-10]}$ for small $\tau$. 
This effect goes in the same direction as that discussed in the
above point 1, but, contrary to it, depends on the adopted geometry (e.g. it is
absent in GH94). The larger dispersion of
the $\alpha_{[2-10]}$ vs. $kT$ relation at high temperature (see Fig. 2) is
due to the increasing importance of anisotropic Compton emission for increasing
$kT$ (see Haardt 1993 for details);\par 
\item{} the contribution of the reflected Compton hump to the emitted
spectrum is $\propto {\rm e}^{-\tau/\mu}$, where $\mu$ is the cosine of the 
inclination angle, so it is smaller for larger $\tau$ 
(for fixed inclination angle), and can be
boosted by anisotropic Compton scattering (HM93) for high temperatures. 
As long as $\tau \gta 0.4$, the contribution of reflection to the [2--10] 
spectrum is marginal ($\Delta \alpha_{[2-10]}\simeq 0.05-0.08$). For lower 
$\tau$ the electron temperature is high enough for anisotropic 
Comptonization to boost reflection. The effect, for fixed $\tau$, is 
larger for larger $\mu$ and larger $kT_{BB}$ (see Fig. 3). Again,
this is an effect whose net result is to flatten the total spectrum as $\tau$
decreases;\par 
\item{} finally we discuss the effect on $\alpha$ of a change in $T_{BB}$. In
fact the Compton $y$ parameter depends not only on  $L_c/L_s$ but also  (only
weakly when $y\lta 1$) on the mean energy of the soft photons. This can be
easily understood considering that the higher the energy of the soft photons
injected in the corona, the lower the number of scatterings needed to reach an
energy $\sim kT$. When $y\ll 1$ the (low) cooling is due only to the first
scattering, so that $L_c/L_s\simeq y$. The relation between $L_c/L_s$ and 
$y$ does not depend on any other parameter. 
But as soon as $y$ increases higher order scatterings
become important in the cooling process. 
The asymptotic limit for $y\gg1$, i.e. when all the photons reached thermal 
equilibrium, is $L_c/L_s=4\Theta/x_0$
where $x_0$ is the dimensionless soft photon mean
energy $\simeq 2.7 kT_{BB}/mc^2$. For this reason, for a given $\tau$ the
equilibrium temperature necessary to sustain a given $L_c/L_s$ 
is (only slightly for $y\simeq 1$) higher (and hence the Compton spectrum
flatter) for higher $T_{BB}$ (for a visualization of how $y$ varies with
$T_{BB}$ see Fig. 1 in Shapiro, Lightman \& Eardly 1976). Large changes in
$T_{BB}$, unlikely to occur in a given source, are required to produce
significant changes in $\alpha$. However, this effect could be relevant when
comparing different classes of sources (e.g. Seyferts and QSOs) where the
difference between the average values of $T_{BB}$  can be large enough to
influence the value of the spectral index of the hard X--rays. As we will
discuss in \S 4, variations of $T_{BB}$ in a particular object, even if small,
can play an important role in the observed variability in the soft X--ray band.
\end{enumerate}

As discussed in \S 2, $\alpha_{[2-10]}$ is a simple least square fit to the
total spectrum, which is not a perfect power law (because of, e.g., anisotropy
and inclusion of reflection). In order to quantify the differences between the
computed spectra and a power law, we associated a fixed percentual error to
each of our 10 computed spectral points. 
By varying the error until a
reduced $\chi^2$ of unity was obtained in the least square fits, we 
estimated a minimum observational precision required to detect the
spectral distortions. For the most distorted spectra
(those at low inclination and low $\tau$) an observational error $\lta$2\%
would result in a reduced $\chi^2$ larger than unity. The typical maximum error
allowed is $\lta$1\%. 

We now discuss the relation between $\alpha_{[2-10]}$ and the X--ray luminosity
$L_c$ on the basis of two alternative limits. The first case corresponds to a
pair dominated corona where the optical depth is determined by the compactness.
In the second case the relation between $\tau$ and $L_c$ is essentially
unknown, and we will consider the limit case in which $\tau$ varies with 
constant luminosity. 

\subsection{The $\alpha-$Flux Relation in a Pair Dominated Corona} 

If the corona is pair dominated, as plausible in the case of
compactnesses $\ell_c\gta 10$, where $\ell_c=(L_c/R)(\sigma_T/m_ec^3)$ and 
$R$ is the typical size--scale of the emitting region 
(see Svensson 1996 for an exact definition of compactness in
different geometries), the optical depth of the hot phase
are determined by the compactness alone. As discussed in HM91, once 
the ratio $L_c/L_s$ and the absolute value of $\ell_c$ 
are specified, there
exists a univocal correspondence of these parameters with $\tau$, under the
assumption that pairs dominate the scattering opacity over normal plasma.
Thus if variations of the compactness are
due to variations of the dissipated luminosity $L_c$, the optical depth and
hence the actual value of $\alpha_{[2-10]}$, depend on the intensity of the
source.\footnote
{Stern et al. (1994) and more recently Poutanen \& Svensson (1996)
and Svensson (1996) pointed out that the exact relation between the rate of
pair production and the heating/cooling rates depends on the details of the
calculations. We have taken the compactness vs. $\tau$ curve given by
Svensson (1996) in order to compute the photon flux in the [2--10] keV band.} 

From the $\alpha_{[2-10]}$ vs. $\tau$ relation computed previously (Fig. 1)
we can therefore derive for a pair dominated corona 
an $\alpha_{[2-10]}$ vs. intensity relation. This is shown in Fig. 4
for two values of the viewing angle. 
It is quite clear (see also Fig. 2 in Stern et al. 1996) that noticeable
variations of $\alpha_{[2-10]}$ (say $\Delta \alpha_{[2-10]} \simeq 0.3$) occur
as the luminosity (and hence the compactness) varies by at least a factor of
20. Thus {\it for a pair dominated corona the spectral 
index must be essentially constant during ``normal" AGN 
X--ray variability (i.e. flux variations within a factor of 2 or so)}. 
Modest spectral steepening can occur during very large intensity increases. 
 
On the other hand a spectral steepening  also implies a decrease of the coronal
temperature (Fig. 2) (see also Pietrini \& Krolik 1995). Since an
observational determination of the latter  with presently available
instrumentation is not easy, we also computed the variation in the hard photon
intensity ([30--100] keV), that would be associated  with a variation in the
[2--10] keV intensity. This is shown in Fig. 5, where  we also report  values
of $\alpha_{[2-10]}$ corresponding to different values of the intensity. The
two intensity scales are arbitrary but their ratio is not.   Due to  the
temperature decrease for increasing luminosity, the intensity in the hard band
varies less than that in the medium band. However the two are positively
correlated. Because of the large overall luminosity variation the steeper and
cooler spectrum does not cross the flatter, hotter one, except at very high
energy. 
 
Different conclusions hold if variations of $\ell_c$ are due to variations of
the linear dimension of the source $R$ rather than to variations of the
intrinsic luminosity. Such a circumstance can occur if the corona is formed by
many active regions rather than being homogeneous. In this case significant
spectral variations could occur without large variations in the output.
Strong effects on the soft X--ray spectrum are expected since the
temperature of the reprocessed radiation scales as $1/R^2$. A detailed
discussion will be presented in the next \S 4. 
 
\subsection{The $\alpha-$Flux Relation in a Low Pair Density Corona} 

If the compactness is $\lta 10$, the corresponding pair optical depth is $\lta
0.1$ (Stern et al. 1994). The main contribution to the
optical depth then comes from "normal" electrons, 
whose amount is not related to the
source luminosity in a simple and easily predictable way. We therefore consider
the case in which the optical depth changes while the dissipated luminosity 
does not vary substantially. 

In Fig. 6 we show the total spectra for some choices of input parameters.
It is evident from the figure that, despite the fact that the total luminosity
in the spectra is the same (when integrated over all the emission angles), the
flux in the [2--10] keV band varies. The spectral index vs. the photon flux in
the [2--10] keV band is shown in Fig. 7. We can see that small observed
variations of the count rate can be accompanied by significant spectral
variations. Another important result, only weakly dependent on details, is that
as long as $\alpha_{[2-10]} \lta 1$, it correlates to the photon flux, i.e. the 
spectrum softens with increasing intensity, while for
steeper spectra the two quantities are anticorrelated. 

From Fig. 6 it is also evident that the flux above 50 keV is going to decrease
for increasing spectral index. The decrement is relatively modest as long as
$\alpha_{[2-10]} \lta 1$ since most of the power is carried by the few high
energy photons, but becomes large as $\alpha_{[2-10]} \gta 1$. Combining the
spectral index intensity relation shown in Fig. 7 with the spectral index
temperature relation we can compute the intensity in the hard X-ray band
([30--100] keV) vs. the [2--10] keV intensity. This is shown in Fig. 8.  For
$\alpha_{[2-10]} \gta 1$ the emission in the [2--10] keV band is  positively
correlated with the emission in the $\gamma-$ray band: small increments in the
X--ray flux are accompanied with larger increments in the $\gamma-$ray flux.
For $\alpha_{[2-10]} \lta 1$ the correlation is in the opposite sense, and is
less strong: small increments in the medium X--ray emission relate to small
decrements of the flux in the $\gamma-$ray band. Around the turning point
$\alpha\simeq 1$ we have the largest variation of the $\gamma$--ray flux for
the smallest variation of the X--ray flux.

In brief, the model predicts that if noticeable spectral variability is
observed while the luminosity variation is less than factor of $\sim 2$,
 then {\it the scattering optical depth of the corona is not (or weakly)
 related to the intrinsic coronal luminosity}. 
This may lead to two possible conclusions: 

\begin{enumerate}
\item{} pairs are {\it not} important as a source of scattering material,\par or
\item{}if pairs dominate the opacity, the linear dimensions of the source
{\it must} vary, i.e. the compactness varies while the luminosity stays 
approximately constant. 
\end{enumerate}

In the latter case the temperature of the soft reprocessed photons must vary,
increasing when the medium energy X--ray spectrum steepens. In any case, 
strong spectral variability can well occur during 
``normal" (within a factor of 2) flux variations. 

 We caution here that in order to measure a variation 
in luminosity
one should measure the intensity up to the high frequency cut off, that is
up to the 100 keV range. In the absence of such information one can {\it
estimate} the luminosity variation from the known  medium energy intensity 
and the expected high energy cut off.

\section{Variability in the Soft X--Ray Domain and Correlations to 
the Medium Band} 

One of the basic features  of the model is  that a fraction of the Comptonized
X-rays is reprocessed into soft thermal photons which are in turn responsible
for the cooling of the corona. This fraction is fixed for fixed geometry,
therefore the model predicts a "perfect"  correlation between the Comptonized
luminosity and the soft reprocessed one. Observationally, the latter component
could be identified with the soft excess observed in Seyfert galaxies, although
there is no general consensus on the very nature of the soft component. 
The soft X--ray emission has been identified as the high energy tail of the UV 
bump (Walter \& Fink 1993), but such claim and in general whether or not 
the soft X--rays are correlated to the UV 
emission is still controversial (see, e.g., Laor et al. 1994 and
Ulrich-Demoulin \& Molendi 1996a, 1996b). The low ROSAT PSPC energy
resolution does not allow to constrain the spectral shape of the excess, and
usually a steep power law is fairly adequate to fit the data (Gondhalekar,
Rouillon--Foley \& Kellet 1996). Higher resolution ASCA data seem to
support a thermal origin of this component (e.g. Leighly et al. 1996), though
absorption features due to highly ionized gas along the AGN line of sight
could be responsible (at least of part) of the excess of residuals in the
single power law fits (e.g. Cappi et al. 1996). Indeed
some recent ASCA data are consistent either with an {\it emission} thermal
component or with {\it absorption} due to highly ionized gas (IC 4329A, Cappi
et al. 1996; NGC 3227, Ptak et. al. 1995). In several objects both a warm
absorber {\it and} a thermal emission component seem to be present (NGC 5548,
Done et al. 1995; NGC 4051, Guainazzi et al. 1995; see also Netzer, Turner \&
George 1994). Mathur, Elvis \& Wilkes (1995) argued that a highly ionized, high
velocity outflowing gas responsible for X--ray and UV absorption might be a
common component in Quasars and should be located outside the Broad Emission
Line Region. 

At least for NGC 5548 there are indications that the soft emission and the
power law vary in a correlated fashion, but the correlation is non linear, i.e.
the soft component varies more (Nandra et al. 1993, Done et al. 1995). In one
case a 30\% increase of the soft counts observed with ROSAT was not accompanied
by a significant variation in the estimated power law component. 
This event suggested that at least soft X--ray ``flares" may well have a 
different origin than reprocessing of hard x--rays (Done et al. 1995). 

We wish to stress here that the behaviour of the observed soft X-ray intensity
depends strongly on the spectrum, i.e. the temperature of the reprocessed
emission $T_{BB}$. The latter depends on the absolute luminosity of the X--ray
emission (we recall that in the model it  equals roughly half of the total
power produced via Comptonization)  and on the size of the radiating surface.
Unfortunately  we do not know whether and how these two quantities are related
with each other. As a first guess we may expect that both $R$ and $L$ scale
with the mass of the source. This may determine a mean value of $L/R^2 $ for a
given source ( sources with smaller masses would have hotter soft components),
however how the ratio varies when the luminosity varies in a given source
remains highly unpredictable. Furthermore, the amplitude of variations in
fixed energy bands can be different  e.g. the hardness ratios can change
substantially when the response of a particular detector is considered.
Comparison with observations can then be made only "filtering" simulated light
curves through the actual response matrix of existing experiments. 

In brief, although there is a one--to--one correspondence between the
Comptonized and reprocessed emission, this holds on global scale, i.e.
considering the integrated total luminosities. What is actually observed as
"count rate" in the given bands can be very different from what is expected on
this simple basis.  In the following we compute the expected behaviour of the
soft X-ray intensity in response to variations of the medium energy intensity
in two limiting cases, for varying luminosity and fixed size of the
reprocessing region and for constant luminosity and varying size. Since most of
the observational data presently available derive from ROSAT we explicitly
compute intensities and hardness ratios in the most commonly used ROSAT bands.

\subsection{Simulated ROSAT light curves} 

We performed simulations of variability patterns according to selected rules
that will be described in the following. The time dependent broad band spectra
were then filtered through a Galactic neutral absorber assuming $N_H = 2\times
10^{20}$ cm$^{-2}$, and finally folded with the ROSAT response matrix. 
We divided
the ROSAT band in three intervals, defining a "hard count rate" as the count
rate in the [0.9--2] keV band ($C_{[0.9-2]}$; it should be kept in mind that in
all the other sections the "hard" band refers to energies $\gta 30$ keV) and a
"soft count rate" as the count rate in the [0.1--0.4] keV band
($C_{[0.1-0.4]}$). We then computed the hardness ratio defined as HR$=
C_{[0.9-2]}/C_{[0.1-0.4]}$. This quantity provides direct spectral information
in the ROSAT range. 

It is important to realize that the dependence of HR on the  temperature of the
soft component is not monotonic. This  is shown in Fig. 9.a for two cases with
regard to the power law component  (both with $0^o$ inclination), one
corresponding to low optical depth $\tau=0.1$ (resulting in a flat spectral
index  $\alpha_{[2-10]}\simeq$0.5-0.6, see Fig. 1), the second to higher
optical depth $\tau=0.65$ (yielding to $\alpha_{[2-10]}\simeq$1.2--1.3) For a
pair dominated source these values of $\tau$ would imply a large difference in
compactness, roughly $10^3$ larger for the high $\tau$ case (see Svensson
1996). Intermediate cases must be comprised between the two curves. The
relation is independent of the actual mechanism causing the variations of
$T_{BB}$, but depends somewhat on the parameters defining the hard power law. 

The behaviour of HR vs. $kT_{BB}$ is easily understandable. For
low temperatures, only the exponential tail of the Planck distribution falls in
the ROSAT band. In this regime a temperature increase leads to a large increase
of the count rate in the soft band only: the spectrum becomes softer. The
softening stops when the peak of the blackbody distribution is within the band,
and the exponential tail starts to affect the intensity in the [0.9--2] keV
band. In the latter regime when the temperature increases the spectrum becomes harder.
 The transition occurs for $kT_{BB}$ around 60 eV. 

 In order to  compute the relation between $T_{BB}$ and the count rate
in the ROSAT band, we have to specify the  mechanisms leading to the temperature
variations. We have considered two different ideal ways in which $T_{BB}$ can
vary, the first corresponding to variations in luminosity for fixed size, the
second to variations in size for fixed luminosity, and discuss them in turn. 

In all the cases discussed in the present section the reprocessed thermal
component is superimposed to a standard multicolor disk emission arising from a
$10^7 M_{\odot}$ black hole radiating at 10\% of the Eddington limit. The
X--ray luminosity is taken as 0.3 of the disk emission and a face--on line of
sight is assumed. We checked that our results are only marginally sensitive to
disk parameters. For rather extreme cases such as a $10^6 M_{\odot}$ black hole
accretion disk radiating at the Eddington limit, with an X--ray luminosity of
only 10\% of the UV disk emission, the ROSAT band is still dominated by the
reprocessed X--rays as long as the temperature of the latter 
component is $\gta
20$ eV. The net effect of increasing the inclination angle is to reduce the
amplitude of variations of the count rate in the soft band. 

\subsection{Variations of Intrinsic Luminosity} 

Suppose the luminosity of the source in the power law component changes (i.e.
$L_c$ changes), while $\tau$ and $\Theta$ remain constant. In this case an
increase in luminosity produces an increase in temperature of the thermal
component, and hence a correlation between medium  and soft X--rays is
expected. 

The hardness ratio plotted in Fig. 9.a can now be plotted as a function of the
intensity which would give rise to the considered temperature variation. This
requires a very large intensity range, a factor $10^7$ for $kT_{BB}$ in the
range 10--316 eV. , implausible for the same source. Thus a single source will
be confined to some restricted interval in intensity, while different sources
could fall in different regions of the plot.  The results are shown in Fig.
9.b. The two curves correspond to the same coronal parameters used in  panel
(a). Clearly the count rate scale is arbitrary, but the absolute value of the
hardness ratio is meaningful, being independent of the source luminosity for a
given spectral shape. For reference, we plot along each curve discrete points 
corresponding to an increase in luminosity by a factor 2, i.e. to an increase
of 20\% in $T_{BB}$ with respect to the preceding one. 
The exact value of $kT_{BB}$ relative to each point can be easily 
read from Fig. 9.a where the same discrete points are marked. 

The count rates in
the two bands are always correlated, though the correlation is not linear. Even
though the temperature only varies as $L^{1/4}$, as long as the temperature is
$\lta 60$ eV the number of soft photons emitted in the ROSAT band changes
significantly for even modest variations of temperature. This regime
corresponds to the decreasing branches of the curves in Fig. 9.b. Along this
branch an increase in luminosity produces an exponential increase of
$C_{[0.1-0.4]}$, while the increase of $C_{[0.9-2]}$ is linear, being dominated
by power law photons. The hardness ratio is clearly anticorrelated to the count
rate. The total count rate shows large variations, being dominated by the
lowest energy channels. 

The behaviour is very different when the peak of the black body spectrum is
well within the ROSAT band, i.e. for $kT\gta 60$ eV. The hardness ratio is
constant as long as the exponential tail of the Planck distribution does not
affect the [0.9--2] band. When this happens, $C_{[0.9-2]}$ responds
exponentially to $\Delta L$, while $C_{[0.1-0.4]}$ varies almost linearly. The
total count rate is almost linear with $\Delta L$, so the spectrum becomes
harder. 

The case of a steep power law component is qualitatively similar but less
extreme. The variations of the hardness ratio are smaller in both branches. 

In a pair dominated source luminosity variations with constant coronal
parameters are, strictly speaking, not allowed. States with different
luminosity  belong to curves with different values of $\tau$. However, since
the separation of the two curves for constant $T_{BB}$ (dotted lines)
corresponds to an increase of $\ell_c$ of a factor $\simeq 10^3$, the shift
corresponding to a factor 2 is clearly negligible. 
   
\subsection{Variations of Linear Dimensions} 

We consider the case of a constant output radiating from a varying  surface
area, while $\tau$ and $\Theta$ remain constant. 
This may represent a case in which the primary hard emission occurs at a
constant rate in an expanding (and/or contracting) corona, or blob, or if the
emission occurs at different time in blobs with roughly the same luminosity but
different sizes. 

Though the luminosity of the hard and soft components is constant, the
temperature of the thermal photons will change giving rise to observable
effects in the soft X--ray domain. To study the problem we varied the linear
size $R$ of the emitting region, so that the resulting black body temperature
covers the range 10--316 eV. Notice that under the assumption
adopted, the intrinsic number of photons emitted scales as $\sqrt R$, while the
temperature as $1/\sqrt R$. Therefore the highest {\it intrinsic} photon
emissivity occurs at the lowest temperatures. 
The relation between hardness ratio and count rates for constant luminosity and
varying area is shown in  Fig. 9.c. The two curves corresponds to the 
same coronal
parameters used in Fig. 9.a and 9.b. The same discrete points plotted in Fig.
9.a and 9.b are also marked, corresponding to a decrease of $R$ of 30\%
(and hence resulting in an increase of $T_{BB}$ of 20\%) with respect
to the preceding one. 

The behaviour of HR 
is similar to that seen in the previous section. As long as $kT_{BB}\lta 60$ 
eV the soft counts increase, while the hard counts are much less 
variable. This regime corresponds to the decreasing branches of the curves 
in Fig. 9.c. Although HR is anticorrelated to the total count 
rate independently of the coronal parameters, the hard and soft count rates 
can be both correlated or anticorrelated, 
depending on whether the hard power law is steep or flat, respectively. 

The behaviour of HR vs. the total count rate is very different when 
$kT_{BB}\gta 60$ eV. Since the total luminosity is constant, 
the total count rate is 
basically constant when the peak 
of the black body distribution is within the ROSAT band. 
This regime corresponds to the almost vertical branches 
of the curves in Fig. 9.c: HR dramatically increases at almost constant 
total count rate. In this regime the peak of the black body 
spectrum moves from the soft band to the hard band. The net result is that 
the two count rates are anticorrelated: as $T_{BB}$ increases $C_{[0.9-2]}$ 
increases while $C_{[0.1-0.4]}$ decreases.

If the corona is pair dominated, variations of $R$ should not happen 
with constant coronal parameters. A reduction in the radius by a factor 
$\simeq 10$ (which roughly corresponds to four successive points counting from
left to right along the curves in Fig. 9.c) 
causes an increase in the opacity and hence a
reduction in the coronal temperature. For $kT_{BB}\lta 60$ eV 
the variations of HR 
will be smaller with respect to cases with constant $\tau$ and $\Theta$. For
larger values of $kT_{BB}$ the situation is similar to what seen before: a
sharp increase of the hardness ratio with a constant count rate. 

In the case of a low pair density corona described in \S 3.2, the coronal 
parameters can vary without any change of size or luminosity 
of the soft component. In terms of ROSAT count rates, this is 
represented by the dashed lines in Fig. 9.c. 
Those lines connect points relative to spectra with different coronal 
parameters, but same $T_{BB}$ and luminosity. If for example an 
increase of $\tau$ occurs, the ROSAT count rate increases as
long as $kT_{BB}\lta 60$ eV, slightly decreases 
for higher black body temperatures (Fig. 9.c). 
The hardness ratio shows a slightly simpler behaviour:
for low temperatures decreases, for $kT_{BB}\gta 30$ eV it increases tending to
become constant at higher temperatures. 

\section{Spectral and Flux Correlations: Comparison with Observations} 

In the following  we compare the results discussed in the present sections 
with some of the
current observations of soft--hard X--ray variability in Seyfert galaxies. 

The two recently studied cases of NGC5548 and Mrk 766 observed with ROSAT
and/or ASCA provide the most detailed available information on spectral and
flux variability of Seyfert I galaxies in the X--rays to date, and also seem to
indicate opposite behaviours. We should warn that 
Mrk 766 is a rather extreme case, not
representative of the Seyfert I class. Indeed this source is classified as a
Narrow Line Seyfert 1 (NLS1). Sources of this subclass tend to show rapid
X--ray variability, unusual steep ROSAT spectra, and strong \ion{Fe}{2}
emission (Pounds \& Brandt 1996). 

In the following we compare those observations with the analysis
developed in the previous sections. We do not aim to "fit" any data, but simply
to interpret (whenever possible) the general variability features in the
context of the reprocessing scenario. 

\subsection{NGC 5548} 

The soft X--ray variability of NGC 5548 observed by ROSAT has been deeply
analyzed by Done et al. (1995). The soft and hard counts seem to go up and down
together, but not in a linear way. In general the source softens (in the ROSAT
band) as it brightens. The hardness ratio roughly doubles (from 0.03 to 0.06)
for a count reduction by a factor 3. Also a state where only the counts in
the ROSAT soft band varied was observed. No simultaneous observations in the
hard X--rays are available. Old GINGA data showed a power law index close to
the canonical value 0.9 plus a reflection hump (Nandra \& Pounds 1994). 

The source  requires a highly variable soft component, and a more steady hard
one. This can be roughly interpreted assuming that variability in the soft band
is due to variations in size of the emitting area, as in the case discussed
 in \S 4.1.2 (see Fig. 9.c). 
The temperature of the thermal component should be less than 60--70 eV, 
not inconsistent with that found 
by Done et al. (1995). 
The intrinsic luminosity of the source should be roughly constant. We may
notice that the lower value of the observed hardness ratio ($0.03$) is lower
that the minimum value consistent with the reprocessing hypothesis 
(see Fig. 9). The discrepancy may be due to the value of $N_H$ for this
source ( $N_H=1.65\times 10^{20}$ cm$^{-2}$; Done et al. (1995) ) which is
lower than 
that used throughout our calculations ($N_H=2\times 10^{20}$ cm$^{-2}$), but  
it may also indicate
that part of the reprocessed thermal flux does not contribute to the cooling of
the corona, or alternatively, as suggested by Done et al. (1995), that there is
another soft component unrelated to the hard one, possibly much broader than 
a single temperature black body. 

\subsection{Mrk 766} 

Mrk 766 has been observed by ROSAT and ASCA,
providing a full coverage of the X--ray domain from 0.1 to 10 keV.  The ROSAT
data has been analyzed by Molendi, Maccacaro \& Schaeidt (1993), Molendi and
Maccacaro (1994) and recently by Leighly et al. (1996) who also analyzed the
ASCA data. ROSAT data show that the source hardens as it brightens, and can be
roughly described by an almost steady soft component plus a variable hard one.
In the ASCA band the source shows evidence of spectral variability above 1 keV
($\Delta \alpha \simeq 0.4$); the steepest state is brighter than the flat
state in the ASCA band by a factor of $\simeq 2$. The power law pivots at an
energy close to 10 keV. If the power law extends up to few hundreds of keV,
then the hard (and weaker in the ASCA band) state is actually more luminous
than the soft state. 

Observations indicate that $\alpha_{[1-10]}$ varies between 0.6 and 1, while
the total luminosity changes by factor of 2 at most. This indicates that either
pairs are not dominant or, if pairs dominate, that the linear size of the
 emitting
region is varying. In the latter case the required change in compactness
is roughly a factor 10. Since the overall X--ray luminosity of the hard
 state is estimated to be  2
times larger than that of the soft state (see below), the region responsible
for the soft  spectrum should be 20 times smaller than that responsible
for the hard  one. The soft state, although fainter by a factor 2, 
should then be characterized by a 
$T_{BB}$ roughly 3--4 times larger than that relative to the hard state. 
Data are consistent with a reduction of the excess normalization by a factor 2
as the power law steepens, but an increase of 3--4 times of $T_{BB}$ in the
soft state is not supported by the data. 

We therefore conclude that the spectral variability in Mrk 766 could be due to
changes of the scattering opacity of the corona without large variations of the
intrinsic luminosity. The opacity of the corona is probably not dominated by
electron-positron pairs. A variation of $\tau$ from $\simeq 0.3-0.4$ to $\simeq
0.1-0.2$ can explain the hardening of the power law, and possibly (part of) the
decrease in the ASCA count rate. We predict that the hard state is
characterized by a coronal temperature $\simeq 150-200$ kev (and hence an
exponential cut-off with an e--folding energy $E_C\simeq 300-400$ keV should be
present), while the soft state should have $kT\simeq 80-100$ keV 
(corresponding to
$E_C\simeq 150-200$ keV). The total X--ray luminosity of the hard state would
then be 2 times larger than that of the soft state. 

Finally we note that the value of hardness ratio in the ROSAT band computed by
Molendi \& Maccacaro (1994) ranges from $\simeq 0.5$ to $\simeq 1$. According
to Fig. 9.a this would imply a temperature of the soft component of $\simeq 130
$ eV, in agreement with the observed hardening with increasing intensity
in the ROSAT band and with the ASCA results. 
  
\section{Summary and Discussion} 

\subsection{Summary}

We have examined the expected  spectral variations within a model involving a
hot corona emitting medium to hard X-rays by Comptonization, coupled to a
cooler optically thick layer which i) provides the input of soft photons for
Comptonization by the hot corona ii) intercepts a fixed fraction of the
Comptonized photons  and reprocesses them into soft thermal photons.  The
fraction of hard photons which are reprocessed and the fraction of soft photons
which are Comptonized have been held fixed to the standard values of 1/2 and 1
respectively which refer to a flattened, sandwich-like geometry. These values
may be somewhat different in different sources but could be constant on "short"
timescales, over which the structure of the emitting region is not expected to
vary  substantially. 

Even with these restrictions the spectral variability is rather complex and
depends on two main issues, the importance of pairs as a source of opacity  and
the  expansion/ contraction of the active regions in the corona possibly
associated with luminosity variations. 

The main point which emerges is that intrinsic spectral variations are governed
by changes in opacity (steeper spectra for larger $\tau$ as long as $\tau \lta
1$). For fixed $L_c/L_s$ this implies that {\it a steepening of $\alpha$ should
correspond to a decrease of the temperature.} This is a robust prediction of
Compton cooled models with fixed geometry and could be tested by the recently
launched XTE and SAX satellites. Should an opposite behaviour be found, the
concept of a quasistationary corona above an accretion disk should be
substantially revised in the sense of a much more chaotic situation. 

The relation between spectral shape and intensity is univocal in the case
of pair dominated coronae. In this case large changes in compactness
(luminosity and /or size) are required for modest spectral variations. These
should appear as either large changes in the 2--10 keV intensity or large
changes in the temperature of the soft reprocessed component.
The case of Mrk 766 (which may be anomalous among Seyfert 1 galaxies)
seems to exclude a pair dominated corona.

In the case of negligible pairs the relation between
luminosity and opacity is essentially unknown. 
Opacity variations could occur with little variations in luminosity and 
give rise to a pivoting of the medium energy spectrum which
can yield little correlation and even anticorrelation between the hard and
medium X-ray fluxes. For constant luminosity and
flat spectral indices ($\alpha_{[2-10]}\lta 1$) 
the hard photon flux varies less and is anticorrelated
with the photon flux in the medium energy band. 

The integrated luminosity in the soft reprocessed  emission is always
proportional to the Comptonized luminosity. However its contribution to
the counts in the band between the galactic absorption cut off and the Carbon
edge depends strongly on the temperature which increases with increasing
luminosity (for fixed size). This can give rise to anomalous observed
behaviours with larger variations in the soft band than in the medium one. 

\subsection{Discussion}

From a more observational point of view, it may be useful to discuss in more
detail what can be deduced about the intrinsic properties of the source from a
given set of "observables" To this end we indicate the [0.1--2] keV band as the
"soft S band", the [2--10] keV band as the "medium M band" and the band above
30 keV as the "hard H band". We define four observable quantities describing
the continuum variability of a source in the X--ray domain: variation of the
[2--10] keV spectral index $\Delta \alpha$, variation of the [0.1--2] keV count
rate $\Delta S$, variation of the [2--10] keV count rate $\Delta M$, and
variation of the hard count rate $\Delta H$. The four quantities defined are in
principle independent, and hence all the possible variability behaviours can be
described  by a combination of possibilities. We will examine the most
interesting ones. 

The main relevant property of an observed X--ray light curve is whether
flux variability in the M band occurs keeping a constant spectral index, or
instead if it is accompanied to spectral variations. The two possibilities,
together with simultaneous observations in the S band, lead to very different
conclusions regarding the very nature of the observed variability. 

As discussed in \S 3 spectral variations in the medium band require a
negligible amount of pairs in the corona, or, for large compactnesses,
variations of the source linear size $R$. Observations in the soft band can
discriminate between the two possibilities, since the latter case would imply
variations of $T_{BB}$ and hence variations of the soft counts. Hence in case
$\Delta S\simeq 0$ the intrinsic luminosity of the X--ray source and the
dimensions of the reprocessing region must not vary; the hard variability in
flux and spectral index is due to variations of optical depth of the corona,
with a negligible contribution of pairs. The model predicts that
$\alpha_{[2-10]}$, whenever steeper than 1, is anticorrelated to the medium
count rate. A positive correlation holds if instead $\alpha_{[2-10]}\lta 1$.
The temperature of the corona decreases for increasing spectral index, and the
flux in the H band is correlated to that in medium band if $\alpha_{[2-10]}
\gta 1$, anticorrelated if $\alpha_{[2-10]} \lta 1$. The case with $\Delta
S\neq 0$ is similar, but the observed variability properties can be due both to
variations of the intrinsic luminosity of the corona, and/or to variations of
the linear size of the reprocessing region. If the corona is pair dominated,
the model predicts that the soft component temperature is correlated to
$\alpha_{[2-10]}$ and anticorrelated to $\Theta$. In case pairs are
unimportant, it is not possible to draw any predictable relation between
$\alpha_{[2-10]}$ and the medium or soft count rates. A full coverage all
through the X--rays to the $\gamma-$rays can determine in principle if the
intrinsic luminosity of the source varies. In case of negligible luminosity
variations, changes in $T_{BB}$ must occur to produce a non null $\Delta S$. On
the other hand, if intrinsic luminosity variations were detected, they alone
could give raise to the observed variability in the soft band without
associated variations of $T_{BB}$. In principle the nature of variability in
the soft band (variations of intrinsic flux or variations of temperature) can
be tested by high spectral resolution observations. \par 

Different conclusions can be drawn if the variability in the medium band occurs
with a constant spectral index. The main result is that the intrinsic
luminosity of the source must vary, but not the properties of the hot corona.
This can happen in a pair dominated source if the variations of compactness are
$\lta 2$. An observation of $\Delta M \gta 10$ with a constant spectral index
would imply that some mechanism different from pair balance works to keep the
coronal optical depth constant. In any case, the model predicts that $\Delta
S\neq 0$. The total counts in the soft and hard bands are positively
correlated, and much stronger variations occurs in the soft band than in the
medium one. $\Theta$ should not vary noticeably, while the flux in H band
follows the variations of the M band. If $\Delta S \simeq 0$ the model can
be ruled out, since any variation of the hard power law with a constant
spectral index must be accompanied by an observable variation in the luminosity
of the soft component. If this does not occur, the soft component cannot be due
to local reprocessing of part of the hard X--rays. 

It should be mentioned that, though the model considered in the paper  
constrains the ratio $L_c/L_s$ to be constant $\simeq 2$, in principle 
different behaviours are possible if instead the ratio $L_c/L_s$ is allowed 
to vary. 
This can occur if the height--to--radius ratio of the scattering 
region(s) is of order unity or larger (see, e.g. , HMG and Stern et al. 1995).
GH94 showed that
in the case of a pair dominated homogeneous plasma with $\Theta \lta 1$, 
an increase of
$L_c/L_s$ (and hence a decrease of $\alpha$) would result in a lower temperature
(and hence a relevant increase in $\tau$ must occur) contrary to the case 
of $\tau$--driven spectral variations with constant $L_c/L_s$.   
Thus, at least in
principle, broad band observations could be used to check the stability of the
Compton--to--soft ratio during spectral variations, and hence the role played 
by reprocessing.  

Finally we wish to mention one aspect of the model neglected so far, namely
time lags between different bands. In the most basic picture of Comptonization
models, the hard X--rays respond to variations of the soft photon flux, and
hence variations in the hard band should follow variations of the soft thermal
photons, exhibiting also a steeper power spectrum (see e.g. the discussion in
Papadakis \& Lawrence 1995). Data in this field are sparse and often
contradictory (see Tagliaferri et al. 1996 and references therein). EXOSAT data
of NGC 4051 seem indicating that more power at high frequency is radiated by
means of high energy photons (Papadakis \& Lawrence 1995), contrary to what
expected from the basic Comptonization theory. 

The above considerations assume that spectral variability is due to variations
of the primary soft photon input. In the model under investigation here however
the soft photons are supposed to be generated by mere reprocessing of a
fraction ($\simeq 50$\%) of the primary hard X--rays. Variations of the
physical conditions occur primarily in the corona. For example, we may expect
that a rapid variation of the heating rate appears in the Comptonized photons
roughly at the same time at every energy, and eventually in the soft component.
The situation appears to be extremely complicated, and we believe that data
interpretation can not avoid to deal with such complications (see, e.g., 
Nowak \& Vaughan 1996). In view of
forthcoming better timing data, in a work in progress we aim to perform a
detailed analysis of the timing properties of an active Comptonizing corona. 

\acknowledgments

FH wishes to thank S. Molendi for stimulating discussions and the Astrophysics
Section of the University of Milan for hospitality. FH and GG thank the ITP in
Santa Barbara where part of this study was done. FH acknowledges financial
support by the Swedish Natural Science Research Council and by the Alice
Wallembergs foundation. This project was also partially supported (FH and GG)
by NSF grant PHY94-07194.

\clearpage

\begin{figure} 
\plotone{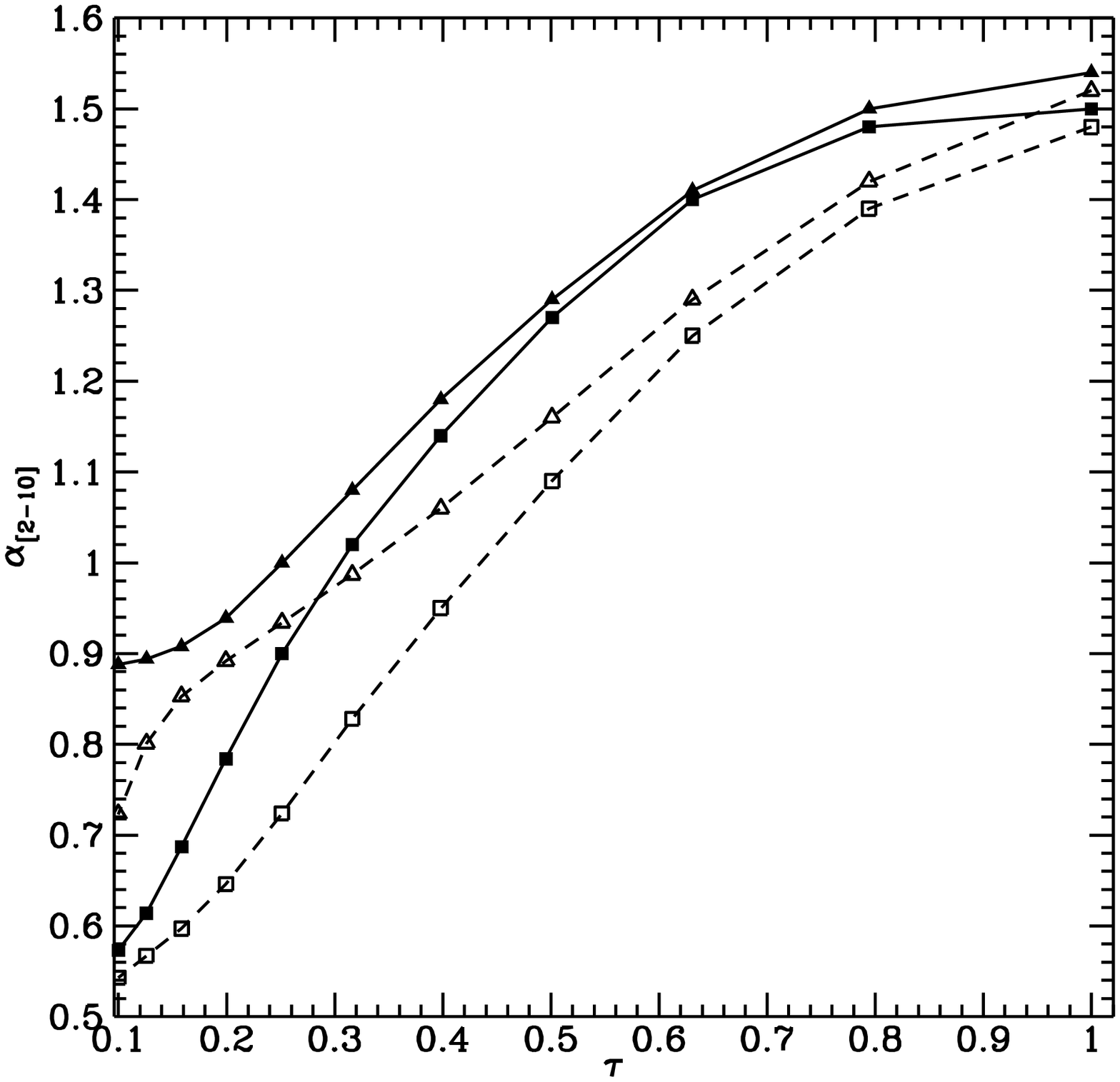}
\caption{The total (direct power law + Comptonized reflection)
spectral index in the [2--10] keV range $\alpha_{[2-10]}$ is plotted against
the coronal scattering optical depth $\tau$. The open symbols connected with
dashed lines refer to spectra computed with $kT_{BB}=100$ eV. The closed
symbols connected with solid lines are for $kT_{BB}=50$ eV. The squares
indicate that a face--on line of sight is assumed, the triangles a 60$^o$
inclination.} 
\end{figure}

\begin{figure}
\plotone{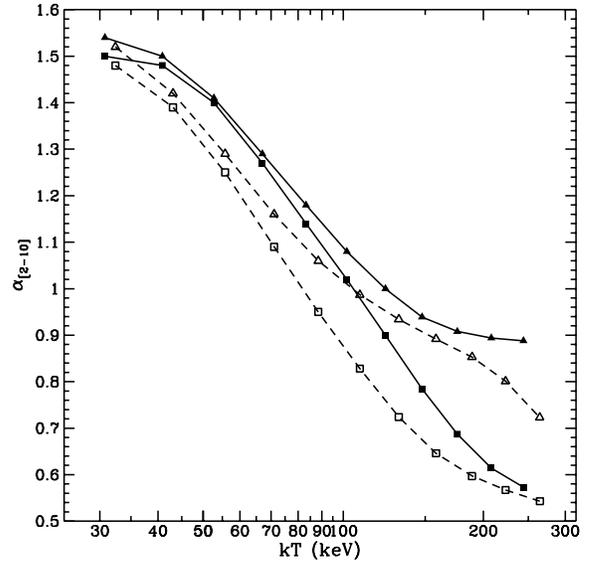}
\caption{The total (direct power law + Comptonized reflection)
spectral index in the [2--10] keV range $\alpha_{[2-10]}$ is plotted against
the coronal electron temperature $kT$. Symbols as in Fig. 1.}
\end{figure}

\begin{figure}
\plotone{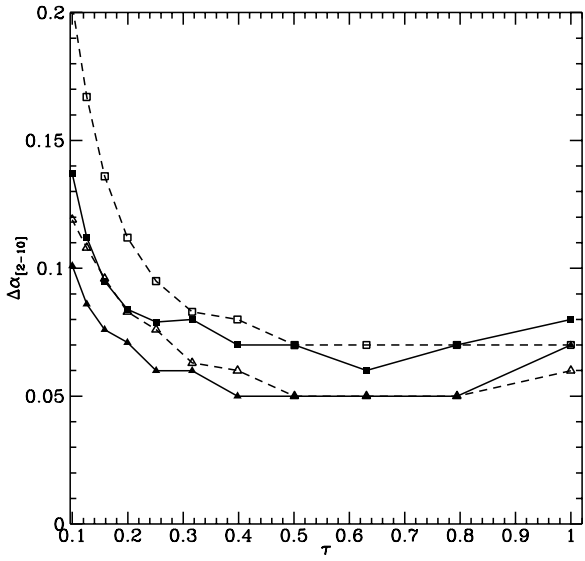}
\caption{The difference between the spectral index of the [2--10] keV intrinsic 
continuum and that of the [2--10] keV total spectrum (continuum+reflection) is 
plotted vs. the optical depth $\tau$. Symbols as in Fig. 1.}
\end{figure}

\begin{figure}
\plotone{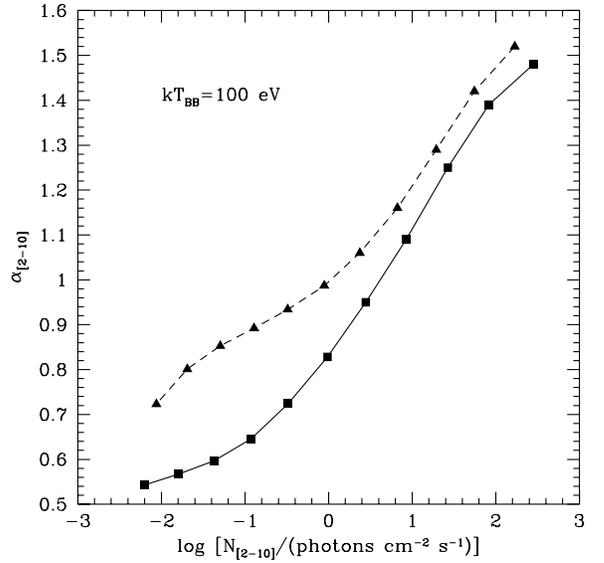} 
\caption{The total spectral index $\alpha_{[2-10]}$ is plotted vs.
the photon flux in the [2--10] keV band (arbitrary normalized) in the case of a
pair dominated corona. Squares connected with solid line refer to a case with a
face--on line of sight, triangles connected by dashed line to an inclination of
60$^o$. $kT_{BB}=100$ eV in both cases.}
\end{figure}

\begin{figure}
\plotone{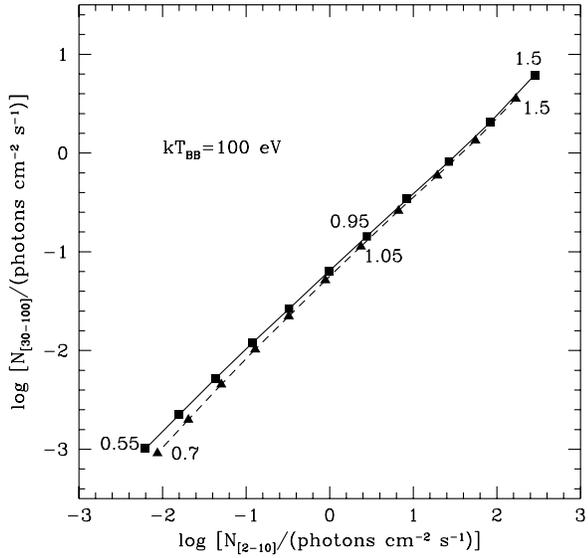} 
\caption{The photon flux in the [30--100] keV band is plotted vs.
the photon flux in the [2--10] keV band in the case of a pair dominated corona.
Normalization is arbitrary. The curves are for $kT_{BB}=100$ eV, and squares
indicate a face--on line of sight, triangles a 60$^o$ inclination. The value of
the total spectral index $\alpha_{[2-10]}$ in the [2--10] keV range is
indicated for some of the spectra.}
\end{figure}

\begin{figure}
\plotone{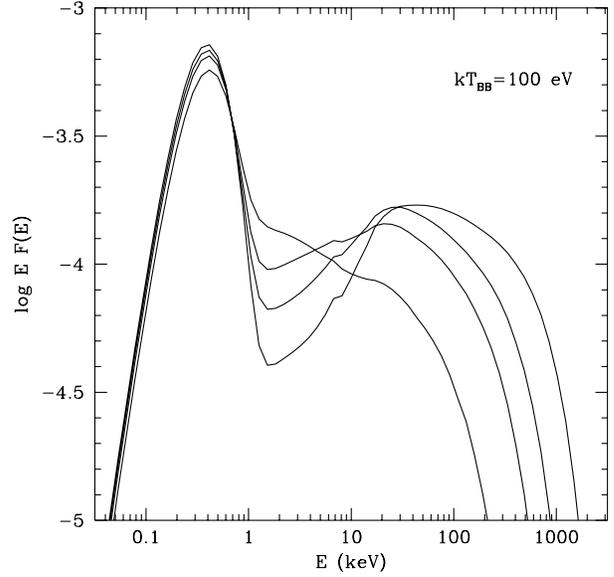} 
\caption{Comptonized spectra for different values of $(\tau,\Theta)$
equilibrium values [(0.63,0.11), (0.32,0.21), (0.2,0.3), (0.1,0.5)]. $kT_{BB}$
is set to 100 eV, and a face--on line of sight is considered.}
\end{figure}

\begin{figure}
\plotone{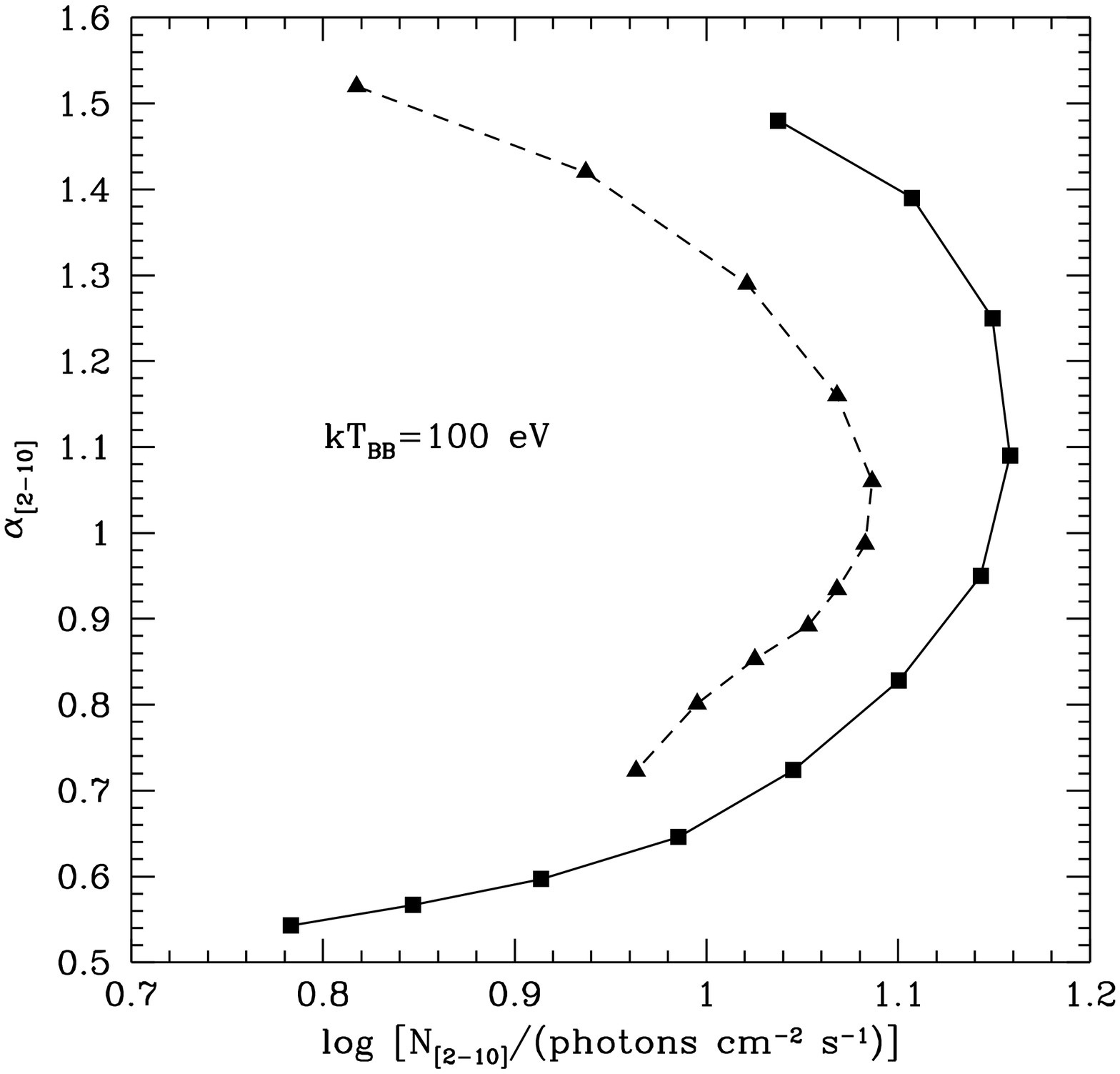} 
\caption{Same as Fig. 4, but in the case of $\tau$ variations with
constant luminosity. Squares with solid line are for a face--on line of sight,
triangles with dashed line for a 60$^o$ inclination.}
\end{figure}

\begin{figure}
\plotone{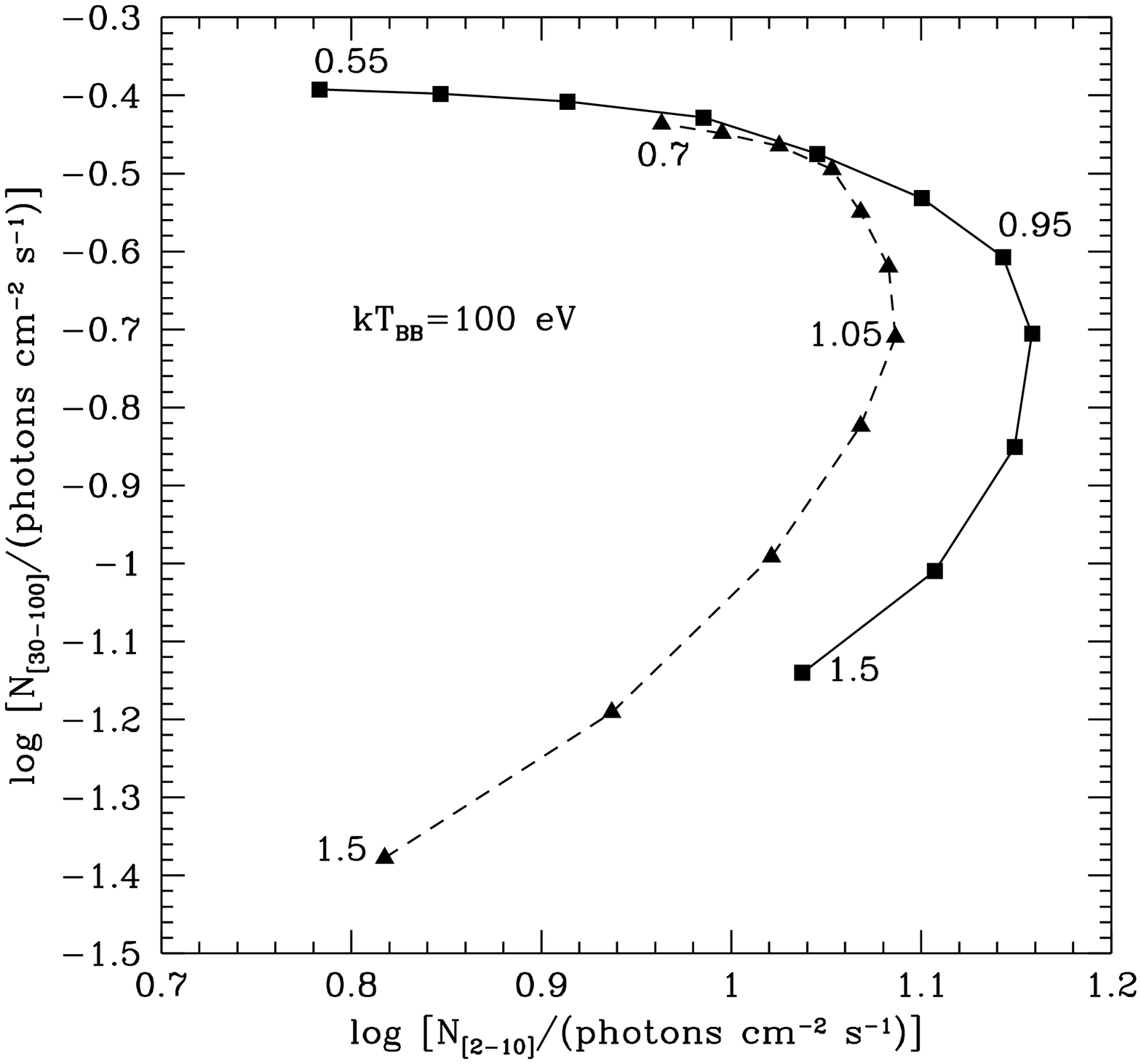} 
\caption{Same as Fig. 5, but in the case of $\tau$ variations with
constant luminosity. Squares with solid line
indicate a face--on line of sight, triangles with dashed line a 60$^o$
inclination.}
\end{figure}

\begin{figure}
\plotone{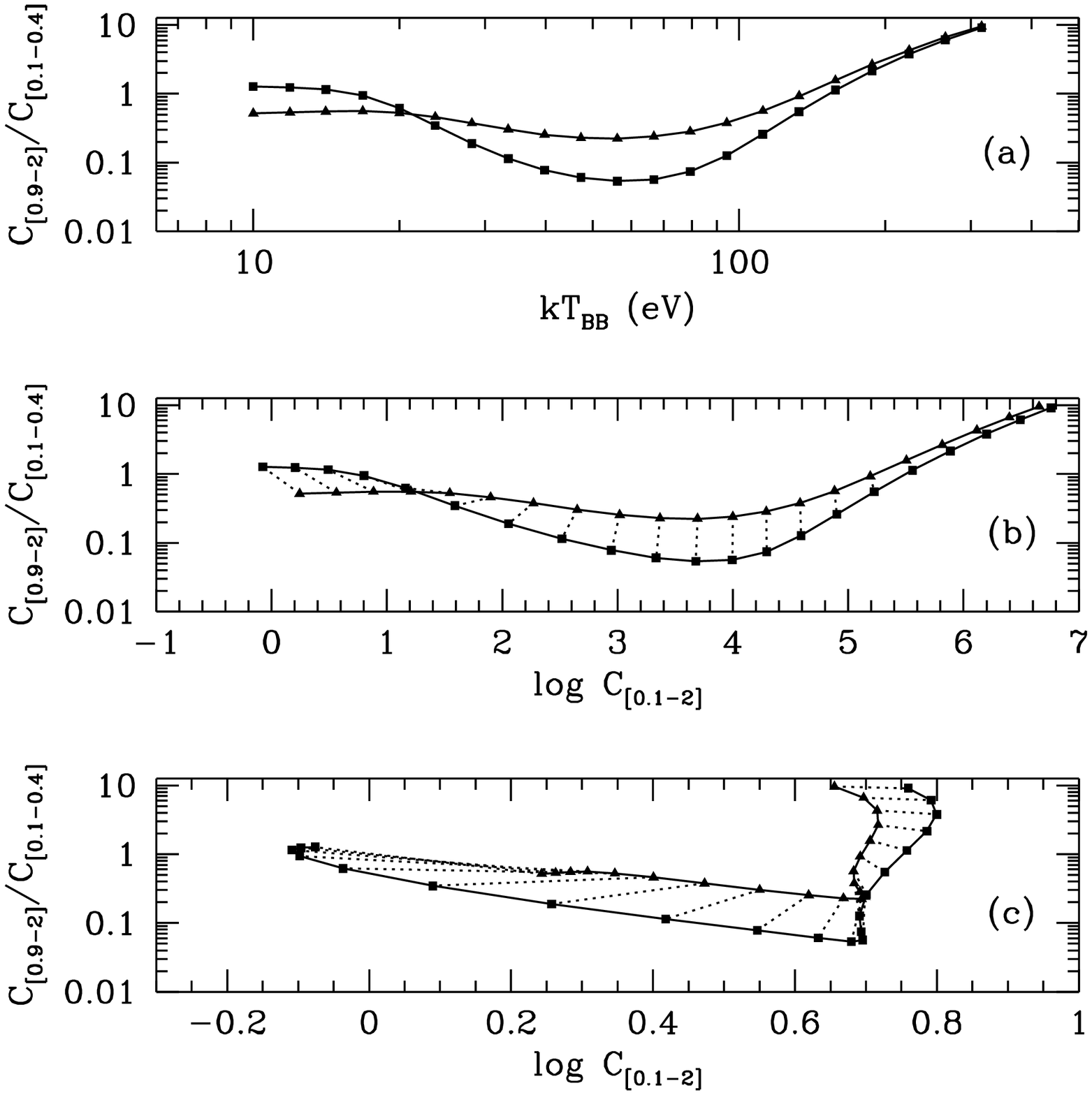} 
\caption{(a) The hardness ratio in the ROSAT band plotted vs. the
temperature of the reprocessed component. Squares refer to spectra computed
with $\tau=0.1$, $\Theta=0.5$, while triangles with $\tau=0.63$, $\Theta=0.11$.
(b) The hardness ratio is plotted vs. the total count rate in the ROSAT band
for a case of pure luminosity variations. Squares and triangles same as (a).
Dashed lines connect spectra having equal $kT_{BB}$. (c) Same as (b) but for
pure variations of the emitting area.}
\end{figure}


\clearpage

\begin{thebibliography}{} 

\bibitem[]{} Arnaud K.A., et al., 1985, \mnras, 217, 105 

\bibitem[]{}
Brandt W.N., Fabian A.C., Nandra K., Reynolds C.S., Brinkmann S., 1994, \mnras,
271, 958 

\bibitem[]{}Burigana C., 1995, \mnras, 272, 481 

\bibitem[]{}
Cappi M., Mihara T., Matsuoka M., Hayashida K., Weaver K.A., Otani C., 1996,
\apj, in press 

\bibitem[]{}
Comastri A., Setti G., Zamorani G., Elvis M., Giommi P., Wilkes B.J., McDowell
J.C., 1991, \apj, 384, 62 

\bibitem[]{}Coppi P.S., Blandford R.D., 1990, \mnras, 245, 453 

\bibitem[]{}Done C., Pounds K.A., Nandra K., Fabian A.C., 1995, \mnras, 
275, 417 


\bibitem[]{}
Elvis M., Wilkes B., McDowell J.C., 1991, in {\it EUV Astronomy}, R. Malina, S.
Bowyer eds. Pergamon Press. New York, p.238 

\bibitem[]{}Fabian A.C., 1994, \apjs, 92, 555 

\bibitem[]{}Ghisellini G., Haardt F., Fabian A.C., 1993, \mnras, 263, L9 

\bibitem[]{}Ghisellini G., Haardt F, 1994, \apj, 429, L53 (GH94) 

\bibitem[]{}Ghisellini G., Haardt F., Matt G., 1994, \mnras, 267, 743 

\bibitem[]{}Gondek D., et al., 1996, \mnras, submitted

\bibitem[]{}Gondhalekar P.M., 
Rouillon-Foley C., Kellet B.J., 1996, \aap, submitted 

\bibitem[]{}
Grandi P., Tagliaferri G., Giommi P., Barr P., Palumbo G.G.C. 1992, \apjs, 82,
93 

\bibitem[]{}Grandi P. Done C., Urry C.M., 1994, \apj, 428, 599 

\bibitem[]{}
Guainazzi M., Mihara T., Otani C., Matsuoka M., 1996, \pasj, in press

\bibitem[]{}Haardt F., 1993, \apj, 413, 680 

\bibitem[]{}Haardt F., 1994, PhD thesis, ISAS/SISSA (Trieste) 

\bibitem[]{}Haardt F., Maraschi L., 1991, \apj, 380, L51 (HM91) 

\bibitem[]{}Haardt F., Maraschi L., 1993, \apj, 413, 507 (HM93) 

\bibitem[]{}Haardt F., Maraschi L., Ghisellini G., 1994, \apj, 432, L95 (HMG) 

\bibitem[]{}Johnson W.N., et al. 1993, \aaps, 97, 21 

\bibitem[]{}
Laor A., Fiore F., Elvis M., Wilkes B.J., Mc Dowell J.C., 1994, \apj, 435, 611 

\bibitem[]{}
Leighly K.M., Mushotzky R.F., Yaqoob T., Kunyeda H., Edelson R., 1996, \apj, in
press 

\bibitem[]{}Liang E.P.T., 1979, \apj, 231, L111 

\bibitem[]{}Madejski G.M., et al., 1995, \apj, 438, 672 

\bibitem[]{}Magdziarz P, Zdziarski A.A, 1995, \mnras, 273, 837 

\bibitem[]{}Maisack M. et al., 1993, \apj, 407, L61 

\bibitem[]{}Mathur S., Elvis M., Wilkes B., 1995, \apj, 452, 230 

\bibitem[]{}Molendi S., Maccacaro T., 1994, \aap, 291, 420 

\bibitem[]{}Molendi S., Maccacaro T., Schaeidt S., 1993, \aap, 271, 18 

\bibitem[]{}Nandra K, et al., 1993, \mnras, 260, 504 

\bibitem[]{}Nandra K, Pounds K.A., 1994, \mnras, 268, 405 

\bibitem[]{}Netzer H., Turner T.J., George I.M., 1994, apj, 435, 106 

\bibitem[]{}Nowak M.A., Vaughan B.A., 1996, \apj, in press

\bibitem[]{}Papadakis I.E., Lawrence A., 1995, \mnras, 272, 161 

\bibitem[]{}Pietrini P., Krolik J.H., 1995, \apj, 447, 526 

\bibitem[]{}
Pounds K.A., Stanger V.J., Turner T.J., King A.R., Czerny N., 1986, \mnras, 224,
443 

\bibitem[]{}
Pounds K.A., Nandra K., Stewart G.C., George I.M., Fabian A.C., 1990, Nat,
344, 132 

\bibitem[]{}
Pounds K.A., Brandt W.N., 1996, in {\it X--Ray Imaging and Spectroscopy of 
Cosmic Hot Plasmas}, Tokyo, in press

\bibitem[]{}Poutanen J., Svensson R., 1996, \apj, in press 

\bibitem[]{}
Poutanen J., Sikora M., Begelman M.C., Magdziarz P., 1996, \apj, in press

\bibitem[]{}
Ptak A., Yaqoob T., Serlemitsos P.J., Mushotzky R., Otani C., 1995, \apj, 436,
L31 

\bibitem[]{}
Saxton R.D., Turner M.J.L., Williams O.R., Stewart G.C., Ohashi T., Kii T.,
1993, \mnras, 262, 63 

\bibitem[]{}Shakura N.I., Sunyaev R.A., 1973, \aap, 24, 337 

\bibitem[]{}Shapiro S.L., Lightman A.P., Eardley D.M., 1976, apj, 204, 187 

\bibitem[]{}
Stern B.E., Poutanen J., Svensson R., Sikora M, Begelman M.C., 1995, \apj, 
449, L13 

\bibitem[]{}Sunyaev R.A., Titarchuk L.G., 1980, \aap, 86, 121 

\bibitem[]{}Svensson R., 1996, \aap, in press 

\bibitem[]{}
Tagliaferri G.,  Bao G., Israel G.L., Stella L., Treves A., 1996, \apj, in press

\bibitem[]{}Titarchuk L.G., Mastichiadis A., 1994, \apj, 433, L33 

\bibitem[]{}Turner T.J., Pounds K.A., 1989, \mnras, 240, 833 

\bibitem[]{}Turner T.J., George I.M., Mushotzky R.F., 1993, \apj, 412, 72 

\bibitem[]{}Ulrich-Demoulin M.-H., Molendi S., 1996a, \apj, 457, 77 

\bibitem[]{}
Ulrich-Demoulin M.-H., Molendi S., 1996b, Wurzburgh Conference contribution, in
press 

\bibitem[]{}Walter R., Fink H.H., 1993, \aap, 274, 105 

\bibitem[]{}Wilkes B.J., Elvis M., 1987, \apj, 323, 243 

\bibitem[]{}
Zdziarski A.A., Lightman A.P., Maciolek-Niedzwiecki A., 1993, \apj, 414, L93 

\bibitem[]{}Zdziarski A.A., et al., 1994, \mnras, 269, L55 

\bibitem[]{}
Zdziarski A.A., Johnson W.N., Done C., Smith D., Mc Naron-Brown K., 1995, \apj,
438, L63 

\bibitem[]{}Zdziarski A.A., Magdziarz P., 1996, \mnras, 279, L21 
 
\end{thebibliography}
\end{document}